\documentclass[aps,preprint]{revtex4} 
\usepackage{graphicx}
\usepackage{amsmath}
\usepackage{float}
\usepackage[font=small,labelfont=bf, figurename=Figure.]{} 

\begin{document}
\title{Domain structure dynamics in the ferromagnetic Kagome-lattice Weyl semimetal Co$_3$Sn$_2$S$_2$ } 

\author{Sandeep Howlader$^1$}

\author{Ranjani Ramachandran$^1$}

\author{Shama$^1$}

\author{Yogesh Singh$^1$}
\author{Goutam Sheet$^1$}
\email{goutam@iisermohali.ac.in}
\affiliation{$^1$Department of Physical Sciences, 
Indian Institute of Science Education and Research Mohali, 
Mohali, Punjab, India}

\begin{abstract}

Co$_3$Sn$_2$S$_2$, a Weyl semimetal that consists of layers of Kagome lattices, transitions from a high temperature paramagnetic phase to a low temperature ferromagnetic phase below 177 K. The phase transition occurs through an intermediate non-trivial magnetic phase, the so called \lq\lq A\rq\rq-phase just below the Curie temperature. The \lq\lq A\rq\rq-phase was earlier linked with a competing anti-ferromagnetic phase, a spin-glass phase and certain indirect measurements indicated the possibility of magnetic Skyrmions in this phase. We have imaged the magnetic domain structure in a single crystal of Co$_3$Sn$_2$S$_2$ at different temperatures, magnetic fields and field-angles by magnetic force microscopy. At low temperatures, we observed stripe domains indicating presence of uniaxial anisotropy. Above 130 K, the domain walls become mobile and they tend to align relatively easily when the magnetic field is increased along the $c$-axis than in the $a-b$ plane. Our detailed study of field-dependent domain dynamics reveal that the anomalous nature of the phase transition just below $T_c$ is dominantly governed by domain wall motion. 
\end{abstract}
\maketitle

Co$_3$Sn$_2$S$_2$ is a magnetic Weyl semimetal and its crystal structure has layers of two-dimensional periodic arrangement of corner sharing triangles -- also called the Kagome lattice \cite{Xu, Vaqueiro, Schnelle, Holder, Kassem1,  Kassem2, Kassem3, Liu}. The systems with Kagome lattice structure are potential hosts of unusual quantum states including quantum spin liquids\cite{Han, Yan}, topological Dirac\cite{Yin,Ye} and Weyl semimetals, Majorana fermions etc. Co$_3$Sn$_2$S$_2$  has in fact shown signatures of Weyl physics where unusually large anomalous Hall conductivity and Hall angle have been measured\cite{Wang}. The coexistence of ferromagnetism and Weyl fermions makes this system special. Moreover, muon spin rotation experiments indicated that the magnetic phase diagram of Co$_3$Sn$_2$S$_2$ is extremely interesting where an antiferromagnetic order is seen to compete with ferromagnetism and there are different scales in terms of temperature and magnetic field where the competition is dominantly in favour of one of the orders\cite{Wang, Liu}. This dynamics gives rise to a non-trivial magnetic phase, the so-called \lq\lq A\rq\rq phase, just below the ferromagnetic transition where puddles of ferromagnetic and antiferromagnetic orders may coexist\cite{Guguchia}. Other time dependent experiments like ac susceptibility also indicated the presence of a non-trivial spin phase just below the ferromagnetic transition\cite{Kassem}. These measurements found evidence of puddles of a frozen spin phase before the entire crystal becomes ferromagnetic. A number of measurements also raised the possibility of a Skyrmionic phase just below the ferromagnetic transition. Earlier it was shown that though Co$_3$Sn$_2$S$_2$ is expected to be half-metallic ferromagnet as per band theory without spin-orbit\cite{Weihrich}, strong-spin orbit coupling causes an intrinsic spin-depolarization effect\cite{Howlader}. Hence, it is also important to consider the role of spin-orbit coupling in understanding the magnetic phase diagram.

In this paper we report direct imaging of the magnetic domains in Co$_3$Sn$_2$S$_2$ by variable temperature magnetic force microscopy under external magnetic fields of different strength applied along different crystallographic directions. The key findings are the observation of stripy domains for which the domain walls become mobile above 130 K. The domain dynamics is anisotropic with respect to the direction of the applied magnetic field. The fraction of the ferromagnetic region exhibiting domains show significant reduction around 160 K, much below the ferromagnetic transition temperature. 

\begin{figure}[h!]
	\centering
		\includegraphics[width=1.0\textwidth]{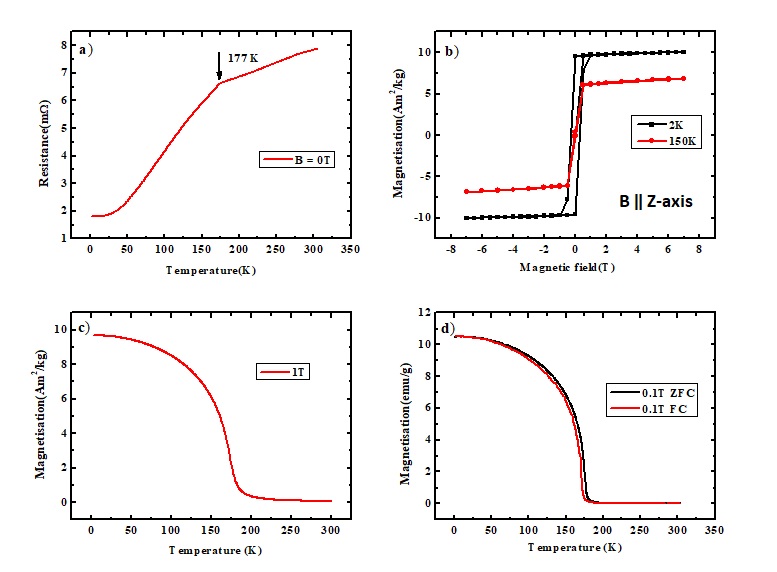}\\
		\caption{(a) Resistivity Vs temperature (b) Field dependent magnetization curves (c),(d) Temperature dependent magnetization curves. }
\end{figure}

High quality single crystals of Co$_3$Sn$_2$S$_2$ were used for all the low temperature microscopic measurements presented here. Co$_3$Sn$_2$S$_2$ single crystals were systhesised by Bridgemann technique\cite{Holder,Kassem1,Kassem2,Kassem3,Dedkov,Umetani}. The crystal structure of crushed single crystals were determined by Reitveld refinement of powder X-ray diffraction. The refinement also confirmed the trigonal crystal structure with space group $R\bar{3}m(166)$\cite{Shama}. The lattice parameters are $a$ = $ b$ = 5.3666 \AA, $ c $ = 13.165 \AA,  and are similar to previously reported values. The stoichiometry of the crystals were confirmed by energy-dispersive X-ray spectroscopy (EDS).\\ 
The bulk electrical and magnetic properties were first measured on the crystals. These measurements were carried out in Quantum Design physical property measurement system(PPMS) equipped with a vibrating sample magnetometer(VSM). Temperature dependence of resistivity is shown in Figure. 1(a) where a sharp kink at 177 K indicates that the Co$_3$Sn$_2$S$_2$ single crystals undergo a phase transition. This is clearly resolved in the temperature dependent magnetization data, where a transition from paramagnetic to ferromagnetic phase is observed (see Figure. 1(c) and Figure.1(d)). The  field dependent magnetization data shown in Figure. 1(b) reveal a narrow hysteresis loop indicating very low coercive field when the field is applied along the $c$-axis of the crystal. This is in contrast to the loop that is obtained when the field is applied along any direction in $a-b$ plane. This indicates that the crystals have uniaxial magnetic anisotropy. The out of plane magnetization attains saturation at a very small magnetic field of 0.1 T.

\begin{figure}[h!]
	\centering
	\includegraphics[width=1\textwidth]{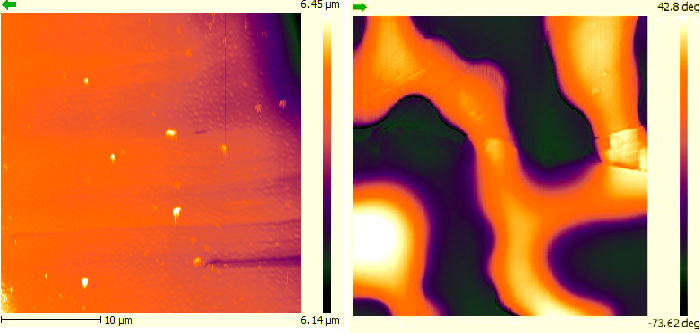}\\
	\caption{(a)Topographic image of Co$_3$Sn$_2$S$_2$ surface and (b) the corresponding MFM phase image at 1.5 K. Arrow on top represents scan direction.}
	
\end{figure}
Now we focus on our low temperature MFM experiments to probe the anomalous \lq\lq A\rq\rq-phase. The MFM experiments were done by a magnetic force microscope (Attocube LT-MFM) working down to 1.5 K and inside a superconducting vector magnet (6T-1T-1T). The MFM scans were done over the sample surface in a dual pass non-contact mode (popularly known as the \lq \lq lift mode\rq \rq). In this mode, during the first pass, the sample is scanned with tip being very close to the sample surface where the Van der waals force dominates. Then the tip is lifted up and another scan is performed while following the topographic profile detected during the first pass. This enables detection of purely magnetic signal free from topographic artifacts. In Figure. 2, we present (a) the topographic and (b) magnetic images obtained on a clean surface of Co$_3$Sn$_2$S$_2$. The stripy contrast appears only in the dual pass phase images and not on the topography indicating that they are the ferromagnetic domains in Co$_3$Sn$_2$S$_2$. To confirm that these stripy features in the dual pass phase image are indeed magnetic domains, we also investigated their behavior with increasing temperature all the way above the Curie temperature. In Figure. 3, we present the MFM images obtained at different temperatures. As it is seen, the stripy contrast regions evolve and smoothly disappear above 177 K, the ferromagnetic Curie temperature of the crystal, confirming that the stripy features are indeed the ferromagnetic domains.
The individual domain width is rather large in this case ($\sim$10 $\mu$m).
 Since the crystals are thick slabs with uniaxial normal anisotropy the stripe domains are naturally expected. The irregular width of domains could be due to presence of defects and other imperfections in the crystal and is expected for such ferromagnetic slabs. 
 \begin{figure}[h!]
	\centering
	\includegraphics[width=1\textwidth]{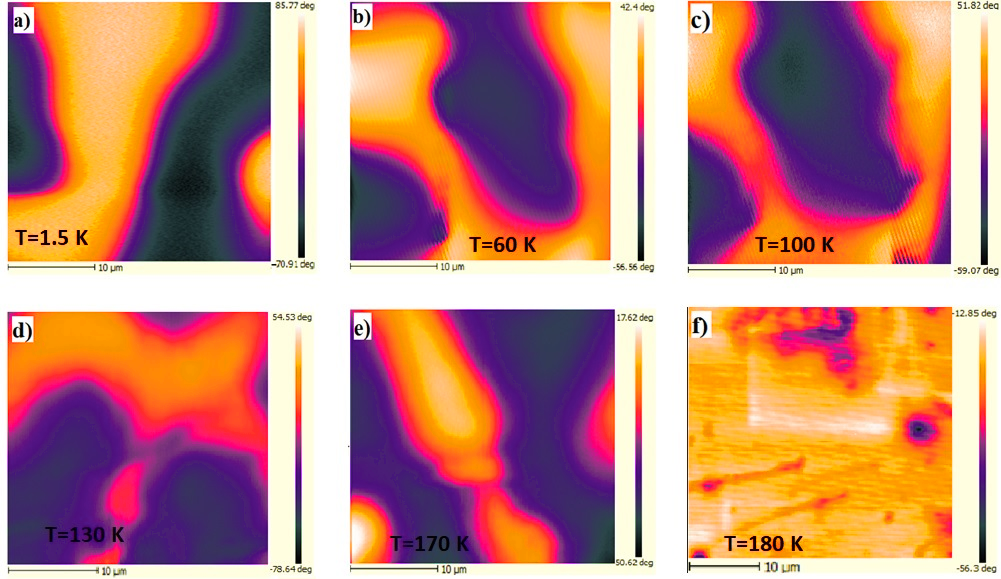}\\
	\caption{ Dual pass phase image at (a) 1.5 K, (b) 60 K, (c) 100 K, (d) 130 K, (e) 170 K and (f) 180 K.}
\end{figure}
\begin{figure}[h!]
	\centering
		\includegraphics[width=1\textwidth]{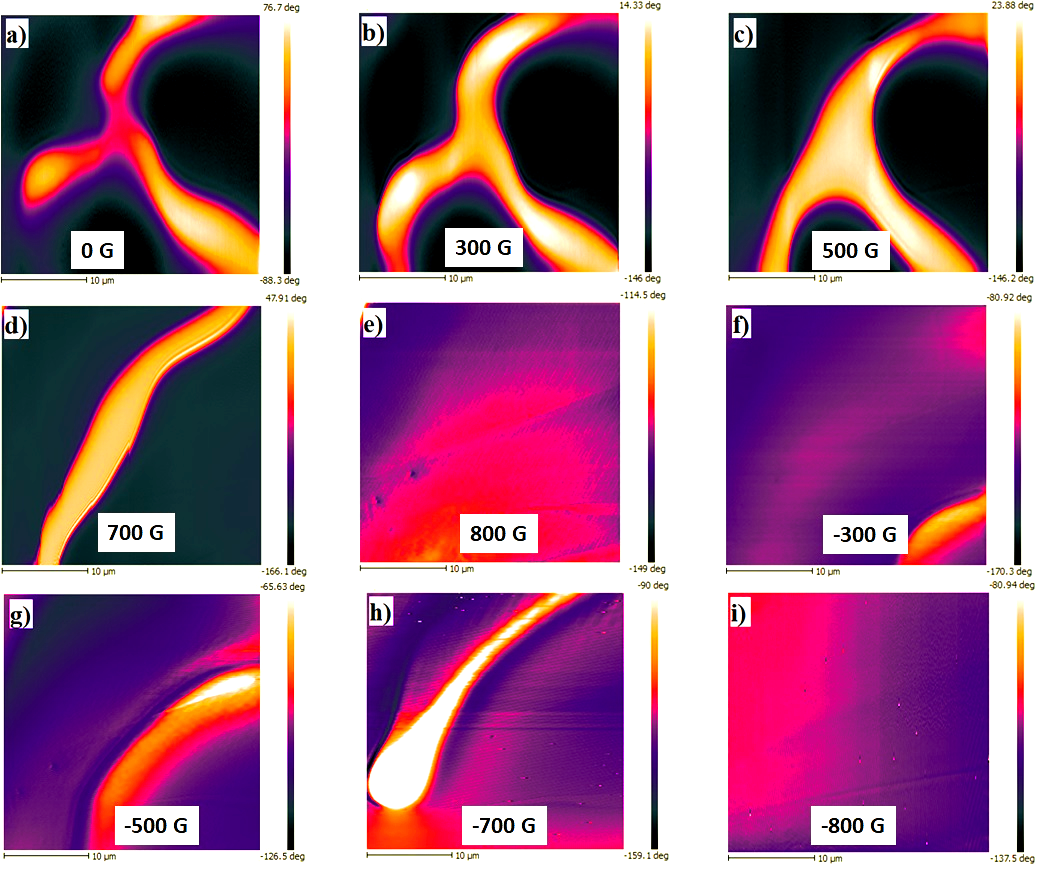}\\
		\caption{Ferromagnetic domains imaged at 1.5 K (a) in absence of applied magnetic field, and in presence of an applied uniaxial magnetic field of (b) 300 G, (c) 500 G, (d) 700 G, (e) 800 G,  (f) -300 G, (g) -500 G, (h) -700 G, and (i) -800 G with field applied along $c$-axis of the crystal.}		
\end{figure}
  The dynamics of the domains were also probed with varying magnetic field. The magnetic field dependent MFM images are presented in Figure 4, where a uniaxial field is applied along the $c$-axis of the crystal. The field was increased/decreased in steps of 100 Gauss in the range 800 Gauss to -800 Gauss. Successive images show expansion of domains polarized along the field direction and  collapse of domains polarized against the field at ~800 Gauss (see Figure. 4(a)-(e)). Furthermore, with application of field in the opposite direction, there is nucleation of domain with \lq \lq negative\rq \rq polarity at -300 G (Figure. 1(f)) is observed which then expand till the  sample is entirely oppositely polarized at -800 G(Figure. 4(g)-(i)).  When the field is applied parallel to the $a-b$ plane of the crystal(Figure. 5),  no noticeable change in the domain structure is observed upto fields of 800 G (Figure 5(a)-(c)). Noticeable alignment of the domain walls along the field direction starts at 1600 G (Figure. 5(f)), however any significant change in the domain structure was not observed till 4000 G (Figure 5(f)-(i)). Therefore it is evident from Figure. 4 and 5 that the saturation field along $a-b$ plane is much larger than along $c$-axis. This is consistent with the results of bulk magnetization measurement indicating that Co$_3$Sn$_2$S$_2$ has a strong uniaxial magnetic anisotropy. The coercive field and saturation field values along different field directions are in good agreement with bulk magnetic measurements.  Hence, it is rational to conclude that the illustrated temperature and the field evolution of the domains is not a local phenomenon but the same dynamics happens throughout the bulk of the crystal.
\begin{figure}[h!]
	\centering
		\includegraphics[width=0.95\textwidth]{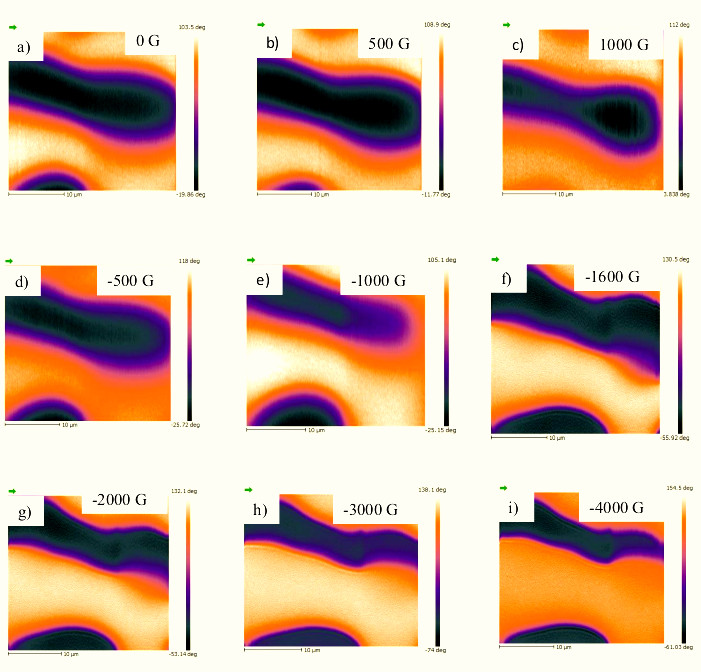}
		\caption{Ferromagnetic domains imaged at 1.5 K (a) in absence of applied magnetic field, and in presence of an applied uniaxial magnetic field  (b) 500 G, (c) 1000 G, (d) -500 G, (e) -1000 G,  (f) -1600 G, (g) -2000 G, (h) -3000 G, and (i) -4000 G along the $a-b$ plane of crystal. Arrow on top represents scan direction.}
\end{figure}    

 \begin{figure}[h!]
	\centering
	\includegraphics[width=.35\textwidth]{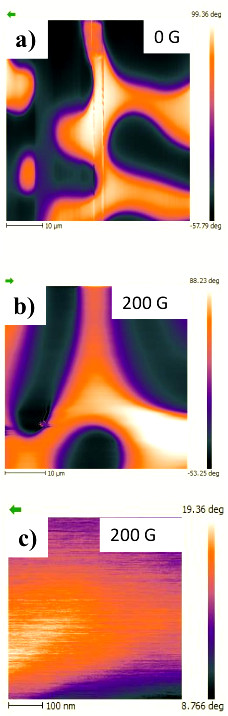}
	\caption{(Domains imaged at 140 K (a) in absence of  magnetic field at 140 K over a scan area of 40 $\mu$m $\times$ 40 $\mu$m , and in presence of 200 G applied along $c$-axis of crystal over a scan area of (b) 40 $\mu$m $\times$ 40 $\mu$m and (c) 500 nm $\times$ 500 nm. Arrow on top represents scan direction.}
	\end{figure}
	
	Previous $\mu$SR and AC susceptibility experiments indicated that the anomalous spin dynamics in the \lq \lq A\rq \rq -phase could be due to emergence of Skyrmions in the \lq \lq A\rq \rq-phase. To probe this possibility we set the temperature at 140 K and applied a magnetic field of 200 Gauss. Under these conditions, the crystal is deep in the \lq \lq A\rq \rq -phase. The MFM image of a 500 nm $\times$ 500 nm area is presented within the limit of the MFM resolution(20 nm lateral), no Skyrmion-like contrast was found(See Figure.6). When the scan was done over a larger area, it was seen that the stripy nature of the domains remained intact even inside the so-called \lq \lq A\rq \rq -phase. Therefore,  based on our MFM experiments role of Skyrmions for the anomalous \lq \lq A\rq \rq -phase can be ruled out. Formation of Skyrmions would cause collapse of the stripy phase that is seen outside the \lq\lq A\rq\rq-phase in the H-T phase diagram. Later, we will also show that dynamics of the stripy pattern with field provide additional support to this claim. \\
     Now it is important to find an alternative explanation of the anomalous spin dynamics of the \lq \lq A\rq \rq -phase. For that, let us revisit the domain dynamics with increasing temperature (Figure 3). As it is seen, with increase in temperature, we also observe a shift in area of scan which is primarily due to thermal drift. The thermal drift between Figure 3(a) and  3(b) is significantly higher than that between Figure 3(b) and 3(c). This is due to the fact that between (a) and (b) there is significantly large difference in temperature than between (b) and (c). From a visual inspection of the set of temperature dependent images, it is clear that increase in temperature ultimately results in increased mobility of domain walls and the domain walls become very highly mobile above 130 K.  The increased mobility of the domain walls at elevated temperatures can be understood from the energetics of the ferromagnetic domains. In anisotropic ferromagnets, the  anisotropy constant is a defining parameter for domain wall energy(E$_{DW}$) and the anisotropy constant can decrease at a much faster rate with temperature than the rate at which the magnetostatic energy (E$_{MS}$) falls with the temperature. This is because E$_{DW} \propto (AK)^{1/2}$ , where $A$-exchange is temperature independent and $K \sim M^p$ where $p$ is an exponent \cite{Jun}. In turn, E$_{MS} \sim M^2$, and $M \sim (T_c-T)^{3/2}$ at low temperatures, while $M \sim (T_c-T)^{0.3-0.5}$ at temperatures close to the ferromagnetic Curie temperature.  Therefore, the ferromagnetic samples will favor the formation of narrower domains along with increased mobility of domain walls with increasing temperature. At the same time the domain width is also expected to decrease with increase in temperature. That is indeed what we observe here (see Figure. 3). Hence,  since Co$_3$Sn$_2$S$_2$ has strong magnetic anisotropy,  the variation of anisotropy constant with temperature alone can account for high mobility of domain walls at higher temperatures. Furthermore, above 160 K, we observe that the net volume fraction of one type of domains decreases significantly even though the temperature is below the Curie temperature. This is also consistent with the recent claim(based on $\mu$SR experiments) that in the \lq \lq A\rq \rq -phase, local anti-ferromagnetic puddles may emerge and coexist with the ferromagnetic domains in Co$_3$Sn$_2$S$_2$.



\begin{figure}[h!]
	\centering
		\includegraphics[width=1\textwidth]{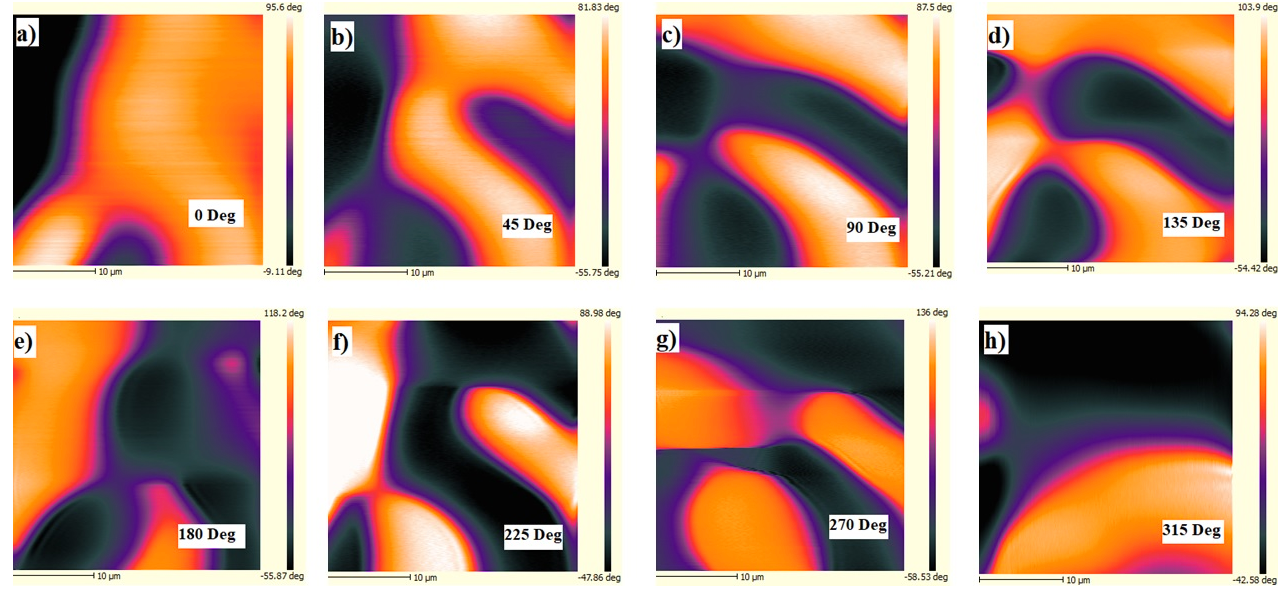}\\
		\caption{Domains imaged in presence of  in-plane rotated magnetic field of 4000 G at 1.5 K }

\end{figure}
\begin{figure}[h!]
	\centering 
	\includegraphics[width=1\textwidth]{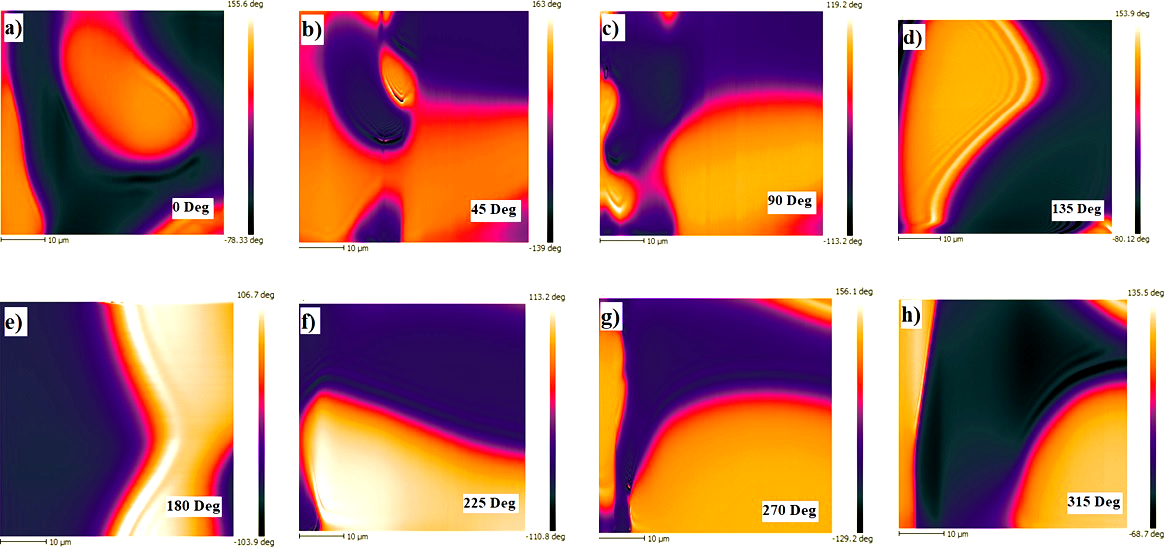}\\
	\caption{Domains imaged in presence of an in-plane rotated magnetic field of 4000 G at 130 K}
	
\end{figure}

 To further probe the relative change in domain wall mobility we performed detailed field-angle dependent experiments at low temperature and within the \lq \lq A\rq \rq- phase. We applied a field of 4 kG and rotated it in the plane of the crystal ($a-b$ plane). It is to be noted here that previous AC susceptibility measurements have indicated that the \lq \lq A\rq \rq -phase has no boundary for magnetic fields applied along the hard axes of Co$_3$Sn$_2$S$_2$ crystals due to strong magnetic anisotropy. To compare the mobility in and out of the \lq\lq A\rq\rq-phase,  the domains were first imaged at 1.5 K and then at significantly higher temperature of 130 K(in the \lq\lq A\rq\rq-phase boundary). We observed that at 1.5 K the domain walls are less mobile which is clearly seen from figure 7. Here, the images look quite similar with minor changes in domains. However, at a significantly higher temperature of 130 K, close to the thermal boundary of \lq \lq A\rq \rq -phase (Figure.8), we observed that the domain walls are extremely mobile. In figure 7, (a) looks significantly different from (b), (c) from (d) and so forth. Here, in some cases, even the magnetic tip facilitated domain wall motion. i.e. the domain walls move along with the cantilever and we obtain a slightly different domain image if we redo imaging at the same area without changing any of the physical parameters such as magnetic field and temperature.


In conclusion, We have probed the \lq \lq A\rq \rq-phase of magnetic phase diagram and presented temperature and field dependent studies of domain structure of ferromagnetic Kagome-lattice semimetal Co$_3$Sn$_2$S$_2$.  We did not detect Skyrmions,in the \lq \lq A\rq \rq -phase, in our MFM experiments. From detailed investigation of temperature and field dependent domain dynamics we surmise that the unusual magnetization dynamics observed in the \lq\lq A\rq\rq-phase might be simply due to enhanced mobility of the domain wall.

We would like to thank Vitalii Vlasko-Vlasov for fruitful discussions. GS acknowledges financial support from a research grant(grant number:\textbf{DST/SJF/PSA-01/2015-16)} under Swarnajayanti Fellowship awarded by the Department of Science and Technology, Govt of India.


\end{document}